\documentclass[aps,prb,twocolumn,groupedaddress,superscriptaddress,nobibnotes]{revtex4-2}
\usepackage{amsmath}
\usepackage{amssymb}
\usepackage{graphicx}
\usepackage{dcolumn}
\usepackage{bm}
\usepackage{color}
\usepackage{subfig}
\usepackage[normalem]{ulem}
\usepackage[none]{hyphenat}
\usepackage{float}
\usepackage{soul}
\usepackage{lipsum}
\usepackage{array}

\newcommand{\PreserveBackslash}[1]{\let\temp=\\#1\let\\=\temp}
\newcolumntype{C}[1]{>{\PreserveBackslash\centering}p{#1}}

\newcommand{\MBT}{MnBi$_2$Te$_4$}

\begin{document}

\author{Mihovil Bosnar}
\email{mihovil.bosnar@gmail.com}
\affiliation{Donostia International Physics Center, 20018 Donostia-San Sebasti\'an, Spain}
\affiliation{Departamento de Pol\'imeros y Materiales Avanzados: F\'isica, Qu\'imica y Tecnolog\'ia, Facultad de Ciencias Qu\'imicas, Universidad del Pa\'is Vasco UPV/EHU, 20018 Donostia-San Sebasti\'an, Spain}

\author{Alexandra Yu.~Vyazovskaya}
\affiliation{Tomsk State University, Tomsk, Russia, 634050}
\affiliation{Saint Petersburg State University, Saint Petersburg, Russia, 199034}

\author{Evgeniy K.~Petrov}
\affiliation{Tomsk State University, Tomsk, Russia, 634050}
\affiliation{Saint Petersburg State University, Saint Petersburg, Russia, 199034}

\author{Evgueni V.~Chulkov}
\affiliation{Donostia International Physics Center, 20018 Donostia-San Sebasti\'an, Spain}
\affiliation{Departamento de Pol\'imeros y Materiales Avanzados: F\'isica, Qu\'imica y Tecnolog\'ia, Facultad de Ciencias Qu\'imicas, Universidad del Pa\'is Vasco UPV/EHU, 20018 Donostia-San Sebasti\'an, Spain}
\affiliation{Saint Petersburg State University, Saint Petersburg, Russia, 199034}
\affiliation{Tomsk State University, Tomsk, Russia, 634050}

\author{Mikhail M.~Otrokov}
\email{mikhail.otrokov@gmail.com}
\affiliation{Centro de Física de Materiales (CFM-MPC), Centro Mixto (CSIC-UPV/EHU), 20018 Donostia-San Sebasti\'an, Spain}
\affiliation{IKERBASQUE, Basque Foundation for Science, 48009 Bilbao, Spain}

\title{High Chern number van der Waals magnetic topological multilayers MnBi$_2$Te$_4$/hBN}

\begin{abstract}
Chern insulators are two-dimensional magnetic topological materials that conduct electricity along their edges via the one-dimensional chiral modes. The number of these modes is a topological invariant called the first Chern number $C$, that defines the quantized Hall conductance as $S_{xy}= C e^2/h$. Increasing $C$ is pivotal for the realization of low-power-consumption topological electronics, but there has been no clear-cut solution of this problem so far, with the majority of existing Chern insulators showing $C=1$. Here, by using state-of-the-art theoretical methods, we propose an efficient approach for the realization of the high-$C$ Chern insulator state in \MBT/hBN van der Waals multilayer heterostructures. We show that a stack of $n$ \MBT{} films with $C=1$ intercalated by hBN monolayers gives rise to a high Chern number state with $C=n$, characterized by $n$ chiral edge modes. This state can be achieved both under the external magnetic field and without it, both cases leading to the quantized Hall conductance $S_{xy}= C e^2/h$. Our results therefore pave way to practical high-$C$ quantized Hall systems.
\end{abstract}

\maketitle

\section{Introduction}

The Chern insulator (CI) state is a quantum phase of two-dimensional (2D) gapped materials with broken time-reversal invariance and nontrivial electronic band topology ~\cite{haldane1988PRL,He.nsr2014, Tokura.nrp2019, Chang.review2022}. It is most straightforwardly probed via Hall measurements, the hallmark being a vanishing longitudinal conductance $S_{xx}$ along with a transversal conductance $S_{xy}$ quantized to integer multiples of the conductance quantum, $Ce^2/h$~\cite{vonKlitzing.prl1980,thouless1982PRL}. Here, $e$ is the electron charge, $h$ is the Planck’s constant, and $C$ is a dimensionless integer called the first Chern number, corresponding to the number of the 1D gapless chiral modes residing at the CI film's edge. The existence of these modes is guaranteed by the nontrivial band topology of CI. 

The edge modes of a CI conduct electricity without dissipation, which could be useful for the construction of novel highly efficient chiral interconnects for low-power-consumption electronics~\cite{Zhang2012ISOP, Zhang.patent2016}. However, the contact resistance between a metal electrode and CI in the envisioned interconnect devices is a bottleneck limiting their performance. To reduce this resistance as much as possible, the number of chiral edge modes, i.e., the Chern number $C$, should be as large as possible~\cite{Zhang2012ISOP,Zhang.patent2016}. Therefore, it is of great interest and importance to engineer CIs with high Chern number.

Historically, the CI state was first observed in 2D electron gases in $1980$ in a transport phenomenon that is now known as quantum Hall effect (QHE) \cite{vonKlitzing.prl1980}. The QHE in this system stems from the formation of Landau levels under the external magnetic field, which drives the system into a topologically-nontrivial state. However, well-defined Landau levels are only possible in systems with high carrier mobility under strong external magnetic fields, which prevents this QHE from a wide applied use. 

Notwithstanding, the developments in the research field of magnetic topological insulators (TIs) in the last decade have allowed a qualitative leap towards QHE without Landau levels. In particular, the quantum anomalous Hall effect (QAHE), a special kind of the QHE that occurs without the external magnetic field, has been observed~\cite{chang2013Science}. It is mainly realized in the thin films of TIs of the (Bi,Sb)$_{2}$Te$_3$ family doped by Cr or/and V atoms ~\cite{chang2013Science, Kou.prl2014, Kandala.ncomms2015, Mogi.apl2015, Gotz.apl2018, Okazaki.nphys2022}. In these systems $C=1$, and although it is theoretically possible to increase $C$ by increasing the dopant concentration and the film thickness~\cite{wang2013PRL}, this has not been experimentally realized to date. 

Instead, complex materials engineering has been resorted to in order to achieve $C=n>1$ state based on the magnetically doped TIs using the following idea. Rather than seeking a high-$C$ state in a particular system, it can be realized by stacking of $n$ CI layers with $C=1$ each. In this case, it is necessary, however, that the adjacent CI layers are efficiently decoupled from each other by a normal, i.e.~topologically trivial, insulator layer. In this way, $C$ up to $4$ and $5$ have been achieved in (Cr,V)$_x$(Bi,Sb)$_{2-x}$Te$_3$/CdSe~\cite{Jiang.cpl2018} and heavily Cr-doped (Bi,Sb)$_{2}$Te$_3$/Cr$_x$(Bi,Sb)$_{2-x}$Te$_3$ multilayers \cite{zhao2020nature}, respectively. 
Although these studies represent a proof-of-concept of the $C$ enhancement approach, it is well known that the potential of magnetically doped TIs for the QAH-based applications is quite limited. Namely, due to a strong disorder in the pnictogen sublattice, which is randomly occupied by Bi, Sb and magnetic dopants, both electronic ~\cite{lee2015pnasu, Chong.nl2020} and magnetic~\cite{lachman2015sciadv} properties of such materials are strongly inhomogeneous. Therefore, the observation of the QAHE in these systems appears to be limited to about $2$ K at best~\cite{Mogi.apl2015, Ou.advmat2018}, with no further improvements achieved over the last several years \cite{Mogi.apl2015}. 

\begin{figure}
    \centering
    \includegraphics{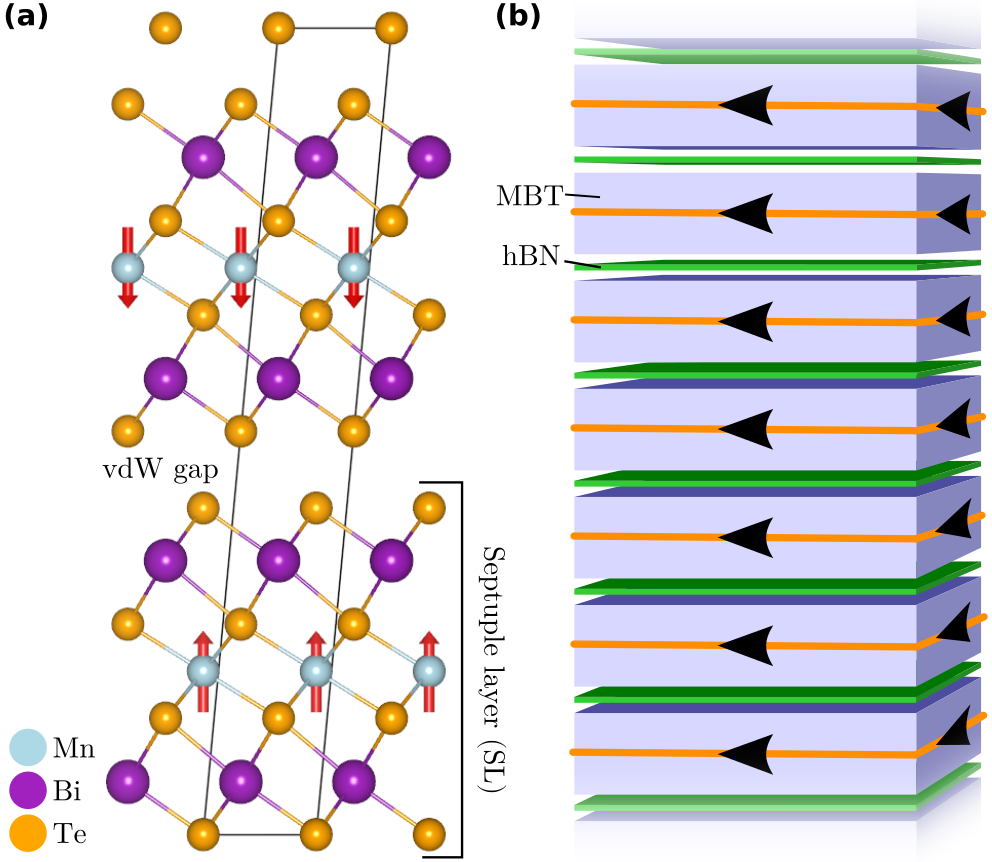}
    \caption{(a) Side view of the bulk \MBT{} (MBT) crystal structure. Red arrows denote Mn local moments. (b) Schematic depiction of the proposed system: MBT films are separated by hBN monolayers to make a vdW multilayer heterostructure with Chern number equal to the number of MBT films, $C=n$. Black arrows depict the direction of the edge currents.}
    \label{fig:bulk_schematic}
\end{figure}

Recently, new systems showing the $C=1$ QAHE have emerged, such as the intrinsic magnetic topological insulators of the \MBT{} (MBT) family \cite{Otrokov2019Nature, Deng.sci2020, Deng.nphys2021}, the twisted bilayer graphene \cite{Serlin.sci2020} and the transition metal dichalcogenide moiré superlattices \cite{Li.nature2021}, opening new opportunities for $C$ engineering. MBT, shown in Fig.~\ref{fig:bulk_schematic}(a), appears as particularly promising due to its van der Waals (vdW) nature, intrinsic combination of nontrivial band topology and long-range antiferromagnetic order ($T_\text{Néel} = 25$ K), as well as large predicted surface band gap \cite{Otrokov2019Nature, Otrokov2019PRL, Mong.prb2010, Li.sciadv2019, Zhang.prl2019, Yan.prm2019, Deng.sci2020, Liu.nmat2020, Ge.nsr2019, Cai.ncomms2022, Hu.prm2021, Liu.ncomms2021, Ying.prb2022, Gao.nat2021, Ovchinnikov.nl2021}. The QAHE in thin MBT flakes has been achieved up to about $1.4$ K, leaving a large room for the observation temperature enhancement. Indeed, a recent study \cite{Ge.nsr2019} demonstrates the actual potential of this material by registering the $C=1$ ($C=2$) QHE up to $30$ K ($13$ K) in its thin flakes, ferromagnetically polarized by external magnetic field. Remarkably, the quantization in this case appears without the Landau levels, in contrast to the conventional QHE observed in 2D electron gas~\cite{vonKlitzing.prl1980}.

Here, inspired by the recent progress on the Q(A)HE in MBT, we propose a novel MBT-based high Chern number material. Namely, we design a multilayer vdW heterostructure, in which thin MBT CI films are stacked on top of each other, interlayed by hexagonal boron nitride (hBN) monolayers that decouple and insulate them from one another (Fig.~\ref{fig:bulk_schematic}(b)). As an inert wide band gap insulator, hBN is an ideal material for such a decoupling, widely used in vdW heterostructure devices as an encapsulating layer or substrate for the stacked 2D materials~\cite{Britnell2012Science,Dean2010NatNano, Geim.nat2013, Novoselov.sci2016}. Using the state-of-the-art density functional theory and tight-binding calculations, we show that the weak vdW bonding between MBT and hBN essentially preserves the $C=1$ CI state in the individual MBT layers. This state can correspond to either (i) the QH insulator phase achieved in thin MBT films under the external magnetic field, but without the Landau levels or (ii) the QAH insulator phase at zero field, if the MBT films are made of the odd number of septuple layer blocks. In either case, stacking $n$ MBT films with $C=1$ interlayed by $(n-1)$ hBN monolayers gives rise to a $C=n$ CI state, with $n$ as large as allowed by the vdW heterostructures growth technology. Our results provide excellent platform for realization of the high-$C$ Chern insulators.

\section{Results}

\subsection{Crystal and electronic structure of MBT/hBN interface}\label{subsec:adsorption_study}

\MBT{} crystallizes in the trigonal $R\bar 3m$-group structure~\cite{Lee.cec2013,Zeugner.cm2019}, made up of septuple layer (SL) blocks, in which hexagonal atomic layers are stacked in the Te-Bi-Te-Mn-Te-Bi-Te sequence, as shown in Fig.~\ref{fig:bulk_schematic}(a). Neighboring SLs are bound by vdW forces. Below $T_\text{Néel} = 25$ K, \MBT{} orders antiferromagnetically due to the antiparallel alignment between the alternating ferromagnetically-ordered Mn layers~\cite{Otrokov2019Nature, Yan.prm2019}, with the local moments pointing out-of-plane (Fig. \ref{fig:bulk_schematic}a).

\begin{table}
 \centering
 \caption{Structural, magnetic, transport and topological characteristics of the MBT$_\mathrm{2SL}$/hBN bilayer for the four adsorption registries: Hollow, Bridge, Top-B, Top-N. $d$ (in \AA{}) is the MBT-hBN interlayer distance (see Fig. 2(a)), $E_{\mathrm{AFM}}$ and $E_{\mathrm{FM}}$ (in meV) are the energies of the respective interlayer magnetic states relative to that of the AFM-hollow case (whose energy is set to zero), $\Delta E_\mathrm{A/F} = E_{\mathrm{AFM}} - E_{\mathrm{FM}}$ (in meV per Mn pair) is the total energy difference of the interlayer AFM and FM states, $S_{xy}$ is the anomalous Hall conductance for the Fermi level lying within the fundamental band gap, and $C$ is the Chern number obtained for the interlayer FM state in MBT$_\mathrm{2SL}$.}
 \begin{tabular}{C{2.2cm}|| C{1.4cm} C{1.4cm} C{1.4cm} C{1.4cm}}
    & Hollow & Bridge & Top-B & Top-N
  \\ \hline
  $d$  & $3.482$ & $3.601$ & $3.573$ & $3.600$
  \\ \hline
  $E_{\mathrm{AFM}}$ & $0.0$ & $5.3$ & $1.6$ & $10.2$ 
  \\ \hline
  $E_{\mathrm{FM}}$ & $1.2$ & $6.4$ & $2.9$ & $11.3$
  \\ \hline
  $\Delta E_\mathrm{A/F}$ & $-1.2$ & $-1.1$ & $-1.3$ & $-1.1$
  \\ \hline
  $S_{xy}$ & $e^2/h$ & $e^2/h$ & $e^2/h$ & $e^2/h$  
  \\ \hline
  $C$ & $1$ & $1$ & $1$ & $1$ 
 \end{tabular}
 \label{tab:bonding_comparison}
\end{table}

\begin{figure*}
\centering
\includegraphics[scale=1.0]{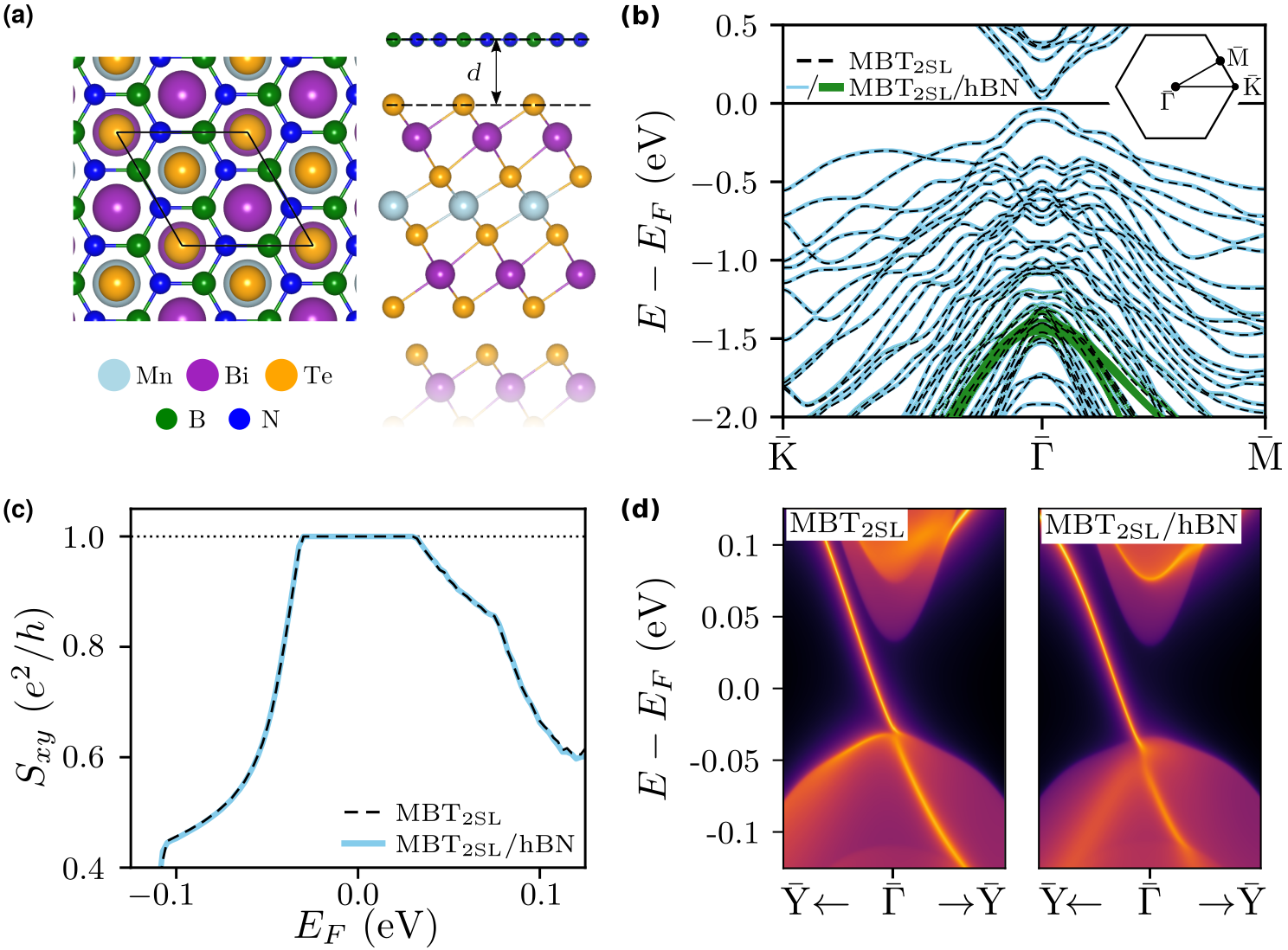}
\caption{(a) Top and side views of the hollow site MBT/hBN adsorption registry. (b) The band structures of MBT$_\text{2SL}$ and MBT$_\text{2SL}$/hBN with hBN contribution shown in green. (c) Fermi energy dependence of the anomalous Hall conductance $S_{xy}(E_F)$ in the units of conductance quantum $e^2/h$ for MBT$_\text{2SL}$ and MBT$_\text{2SL}$/hBN. $E_F=0$ corresponds to the center of the band gap. (d) The edge electronic band structures for MBT$_\text{2SL}$ (left) and MBT$_\text{2SL}$/hBN (right). The regions with a continuous spectrum correspond to the 2D bulk states projected onto the 1D Brillouin zone. The edge crystal structure is shown in Supplementary Figure S7. The data in (b-d) were calculated for the FM interlayer alignment in MBT$_\text{2SL}$ (see text).}
\label{fig:bilayer_overview}
\end{figure*}

We start our study from the consideration of structural, magnetic and electronic properties of the MBT/hBN bilayer made of the $2$ SL thick MBT film (MBT$_{\mathrm{2SL}}$) and hBN monolayer. This can be considered a minimal system because it contains all of the essential characteristics of MBT, such as intra- and interlayer exchange couplings as well as the non-trivial topology in the forced FM state \cite{Otrokov2019PRL}, so it can be used to test whether they are affected by hBN.

The MBT and hBN basal planes are symmetry compatible and show a good lattice parameter matching in the MBT($1\times1$)/hBN($\sqrt{3}\times\sqrt{3}$) configuration, with a mismatch of only about $0.6$ \%. The optimal hBN adsorption geometry was then determined by the comparison of total energies of structurally optimized MBT$_\text{2SL}$/hBN in four high-symmetry registries of such configuration, shown in Supplementary Fig.~S3(a)-(d). These energies were calculated starting from both interlayer FM and AFM spin configurations of MBT$_\text{2SL}$, assuming the out-of-plane magnetic moment direction, to determine a possible influence of hBN on the interlayer exchange coupling, which is in MBT significantly weaker than the intralayer one (the latter is addressed below, as well). For more details, see Supplementary Note~SII A 1.

The relevant numerical results are listed in Table.~\ref{tab:bonding_comparison}. It can be seen that for all adsorption registries the interlayer distance $d$ between MBT and hBN is about $3.5-3.6$ \AA{}, and the AFM spin configuration is lower than the FM configuration by at least $\Delta E_\mathrm{A/F} = E_{\mathrm{AFM}} - E_{\mathrm{FM}} \simeq 1\ \mathrm{meV}$ per Mn pair. The lowest energy is obtained for the hollow site geometry, shown in Fig.~\ref{fig:bilayer_overview}(a) and Supplementary Fig.~S3(a). The large $d$ on one hand and the overall similarity of the energy differences $\Delta E_\mathrm{A/F}$ in MBT$_\text{2SL}$/hBN and MBT$_\text{2SL}$~\cite{Otrokov2019PRL} on the other suggest the vdW bonding of MBT and hBN.

Having determined the MBT/hBN interface geometry, we can explore the effect of hBN on the MBT$_\text{2SL}$ electronic structure and topology. Since we are interested in the CI state, the FM interlayer spin alignment is considered here, for which MBT$_\text{2SL}$ has been predicted to have $C=1$ \cite{Otrokov2019PRL}. Noteworthy, in experiments the FM alignment in MBT is achieved by the external magnetic field application \cite{Lee.prr2019, Deng.sci2020, Liu.nmat2020, Ge.nsr2019}. 

Comparison of the band structure along the $\overline{\mathrm{K}}-\overline{\Gamma}-\overline{\mathrm{M}}$ path for MBT$_\text{2SL}$/hBN in the hollow site registry and pure MBT$_\text{2SL}$ is shown in Fig.~\ref{fig:bilayer_overview}(b). Their similarity near the Fermi level is immediately obvious and the band gaps, $63.5$ and $62.7$ meV, respectively, are very close. The hBN states can be found about $1.5\ \mathrm{eV}$ below the Fermi level (and deeper) as well as over $3.2\ \mathrm{eV}$ above it. Furthermore, Fig.~\ref{fig:bilayer_overview}(c) shows that the Fermi level dependencies of the anomalous Hall conductance, $S_{xy}(E_F)$, of MBT$_\text{2SL}$/hBN and MBT$_\text{2SL}$ are very well matched, too. In particular, in both cases $S_{xy}$ is constant inside the band gap, where the actual Fermi level is, and equal to one conductance quantum $e^2/h$, indicating the CI state with $C=1$. Accordingly, the edge spectral function of MBT$_\text{2SL}$/hBN in Fig.~\ref{fig:bilayer_overview}(d) features a single chiral edge mode traversing the band gap, similar to  MBT$_\text{2SL}$. Finally, a Wilson loop method calculation for MBT$_\text{2SL}$/hBN yields $|C|=1$ as well, with the sign depending on the magnetization direction, as expected for a CI~\cite{Chang.review2022}. 

The calculations of the band structures, $S_{xy}(E_F)$ and $C$ for the other three high-symmetry registries show that the same results hold for all of them (see~Table~\ref{tab:bonding_comparison} and the Supplementary Fig.~S3(e)-(k) for visualization). These findings clearly confirm the vdW nature of bond between hBN and MBT, which preserves the CI state in the forced FM phase of the MBT$_\text{2SL}$ film. 

We note that recently it has been reported elsewhere~\cite{Gao2021PRM} that while hBN and MBT are bound by vdW interaction, the interlayer FM configuration in MBT$_{\mathrm{2SL}}$/hBN becomes significantly lower in energy than the AFM one (by up to $45$ meV) for any adsorption registry. We have attempted to reproduce those results by retracing the steps outlined in Ref.~\cite{Gao2021PRM} (see Supplementary Note~SII A 2), but arrived to the same results that we present here. We believe that our result is correct, on the physical basis that the vdW interaction along with the insulating character of hBN should not produce such a drastic effect on magnetism as it was found in Ref.~\cite{Gao2021PRM}.

Finally, we confirm the implicit assumptions of preference for the FM intralayer spin configuration and the out-of-plane easy axis direction by total energy calculations on MBT$_\text{1SL}$/hBN (see Supplementary Notes SII A 3 and SII A 4, respectively). The former calculation reveals that the FM configuration is by $16.9$ meV (per Mn pair) lower than the AFM configuration while the latter yields a positive magnetic anisotropy energy of $0.07\ \mathrm{meV}$ per Mn atom (vs. $0.074$ meV in pure MBT$_\text{1SL}$), meaning that the easy axis indeed stays out-of-plane. Thus, neither the intralayer magnetic order of MBT nor its magnetic anisotropy are not changed by interfacing with hBN. The above results concerning the insensitivity of magnetic, electronic and topological properties of MBT to hBN should hold for thicker MBT films as well because of the vdW nature of the bond. Thus, we conclude that hBN can be efficiently used to decouple MBT CI layers from each other, without altering their properties.

\subsection{High-$C$ state in the forced FM phase}\label{subsec:bilayer_heterostructres}

We can now proceed with the study of the topological properties of the MBT/hBN multilayer heterostructures. We first note that in experiments the CI state in the forced FM phase, achieved through the application of the external magnetic field, is observed in thin MBT flakes made of both even and odd number of SLs \cite{Deng.sci2020, Liu.nmat2020, Ge.nsr2019, Cai.ncomms2022, Bai.arxiv2022, Hu.prm2021, Liu.ncomms2021, Ying.prb2022, lujan2022:natcomm, Gao.nat2021, Ovchinnikov.nl2021}. According to the previous density functional theory calculations \cite{Otrokov2019PRL}, the minimal MBT film thickness required for the realization of such a $C=1$ state is two SL blocks. Let us therefore consider $n$MBT$_\text{2SL}$/hBN, $n>1$, heterostructures, in which $n$ films of MBT$_\text{2SL}$ are interlayed by $(n-1)$ hBN monolayers, as schematically depicted in Fig.~\ref{fig:bulk_schematic}(b). We will assume that all MBT$_\text{2SL}$ films are FM-polarized in the $+z$ direction by the external magnetic field. These heterostructures were constructed based on the structure of the MBT/hBN/MBT system that was determined after a series of calculations outlined in the Supplementary Note~SII B 1. 

Fig.~\ref{fig:multilayer_bands} shows the calculated low-energy band structures along the $\overline{\mathrm{K}}-\overline{\Gamma}-\overline{\mathrm{M}}$ path for $n$MBT$_\text{2SL}$/hBN with $n=2, 3, 4$ and $5$. The band structures basically correspond to that of a free-standing MBT$_\text{2SL}$ repeated $n$ times, slightly shifted in energy due to a slight variation of the electrostatic potential across the multilayer. The bands stemming from the spatial inversion equivalent MBT$_{\mathrm{2SL}}$ layers come in pairs, as it is seen in insets to Fig.~\ref{fig:multilayer_bands}. In the corresponding $S_{xy}(E_F)$ dependencies, shown in Fig.~\ref{fig:edge_ldos_c2_c3}(a), there are plateaus in the band gap that are equal to an integer number of conductance quanta, the integer being equal to $n$, suggesting $C=n$ state in the respective multilayers. Accordingly, in the plots of the calculated edge spectral functions, shown in Fig.~\ref{fig:edge_ldos_c2_c3}(b), two (three) edge modes can be seen for the $n=2\ (n=3)$ system.

\begin{figure*}
    \centering
    \includegraphics{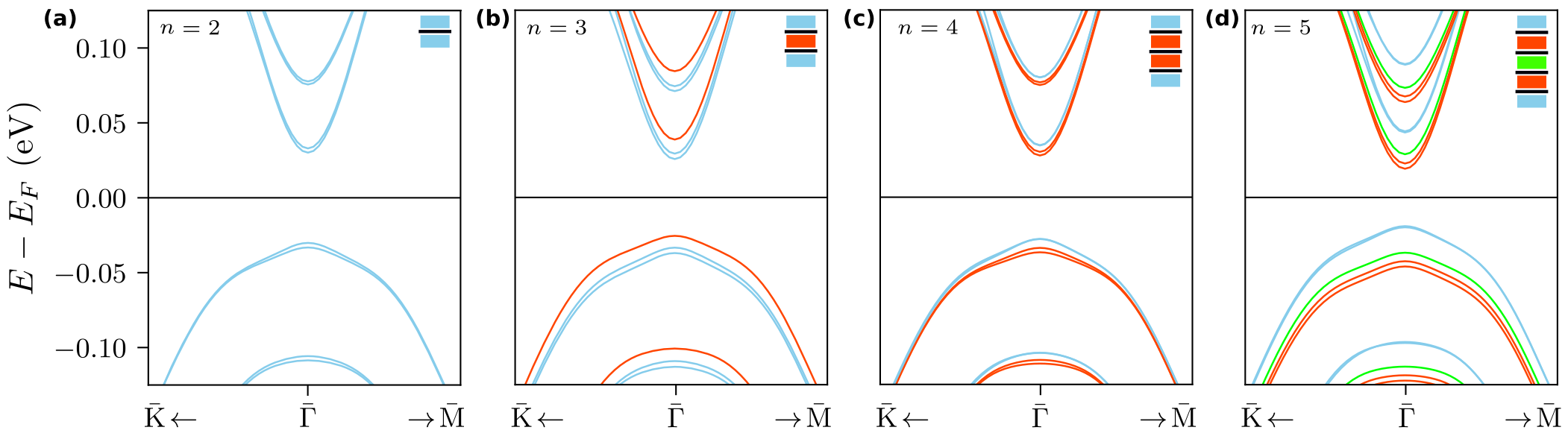}
    \caption{Electronic band structures calculated along the $\overline{\mathrm{K}}-\overline{\Gamma}-\overline{\mathrm{M}}$ path in the 2D Brillouin zone of the $n$MBT$_{\mathrm{2SL}}$/hBN multilayers for (a) $n=2$, (b) $n=3$, (c) $n=4$ and (d) $n=5$. In the insets, the colored blocks depict the equivalent MBT$_{\mathrm{2SL}}$ layers from which the bands of corresponding color dominantly stem. The black lines between the blocks represent hBN layers that separate them.}
    \label{fig:multilayer_bands}
\end{figure*}

\begin{figure*}
    \centering
    \includegraphics{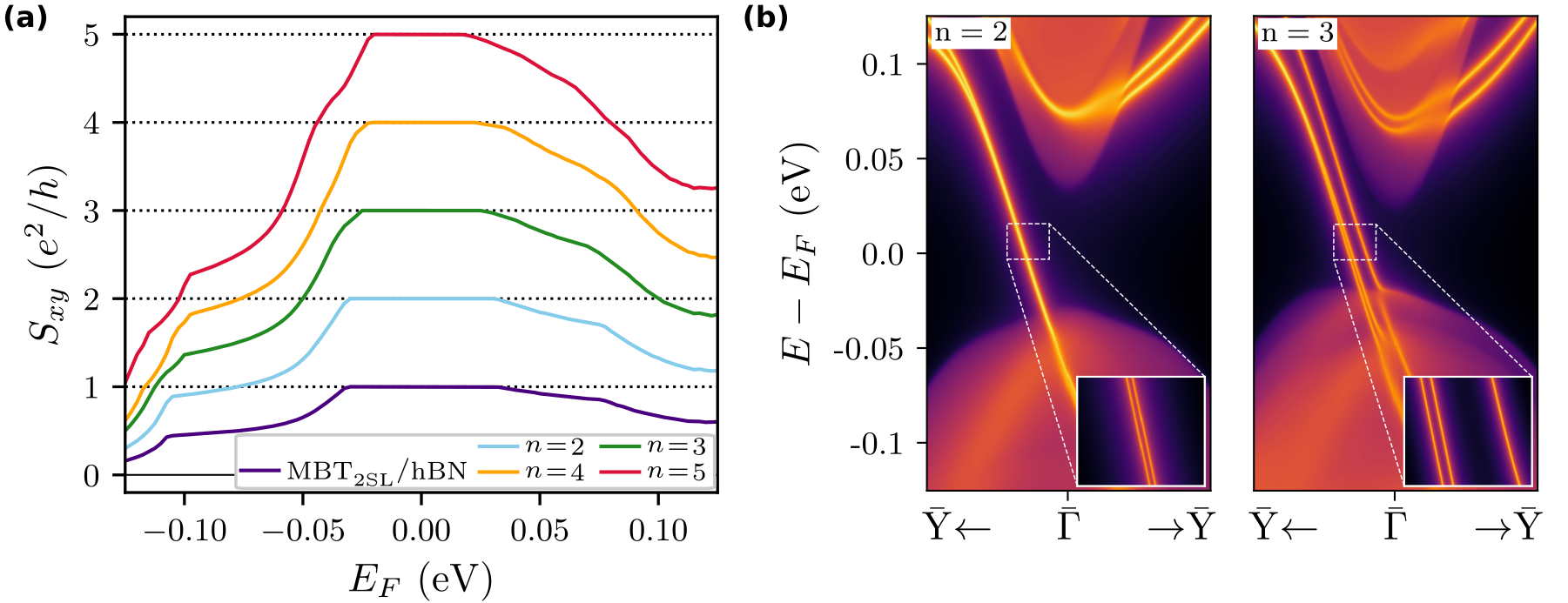}
    \caption{The Fermi energy dependence of anomalous Hall conductance $S_{xy}(E_F)$ in the units of conductance quantum $e^2/h$ for MBT$_{\mathrm{2SL}}$/hBN and $n$MBT$_{\mathrm{2SL}}$/hBN, $n=2,\dots,5$. $E_F=0$ corresponds to the center of the band gap in each case. (b) The edge electronic structure of $n$MBT$_{\mathrm{2SL}}$/hBN for $n=2$ and $n=3$.}
    \label{fig:edge_ldos_c2_c3}
\end{figure*}

From these results we can conclude that $C=n$ in the examined heterostructures and infer by induction that the same will hold for heterostructures with greater $n$.

\subsection{High-$C$ QAH state: odd number of SLs}\label{subsec:trilayer_heterostructures}

\begin{figure*}
    \centering
    \includegraphics{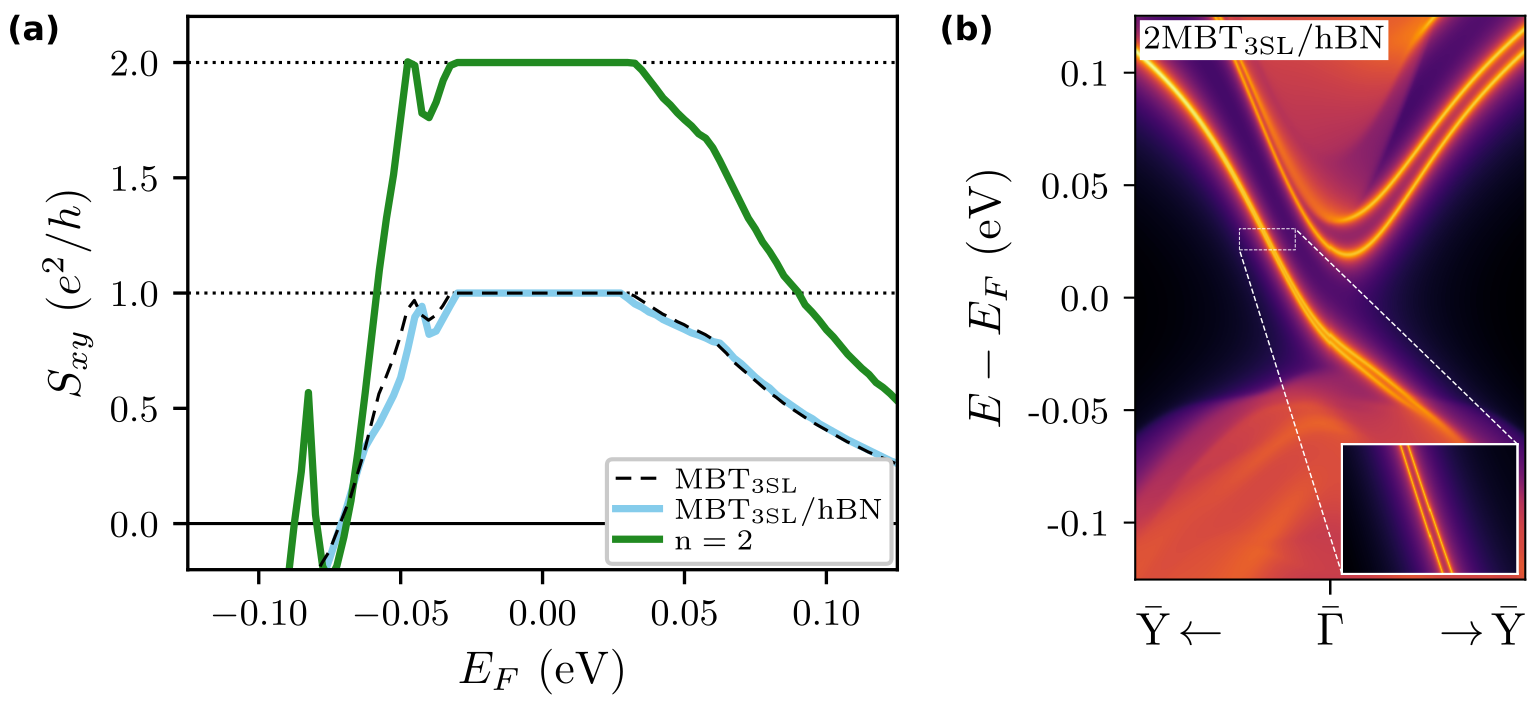}
    \caption{(a) The Fermi energy dependence of the anomalous Hall conductance $S_{xy}(E_F)$ in the units of conductance quantum $e^2/h$ for $\mathrm{MBT}_{\mathrm{3SL}}$, $\mathrm{MBT}_{\mathrm{3SL}}/\mathrm{hBN}$ and  $2\mathrm{MBT}_{\mathrm{3SL}}/\mathrm{hBN}$. $E_F=0$ corresponds to the center of the band gap in each case. (b) The edge electronic structure of $2$MBT$_{3\mathrm{SL}}$/hBN.}
    \label{fig:trilayer_ahc}
\end{figure*}

In the thin films made of an odd number of SLs, MBT has been predicted \cite{Otrokov2019PRL, Li.sciadv2019} and subsequently experimentally confirmed \cite{Deng.sci2020} to show the intrinsic $C=1$ QAHE. We therefore now turn to the topological properties of the MBT/hBN heterostructures based on the thinnest possible odd-SL MBT film, i.e. MBT$_\text{3SL}$ \cite{Otrokov2019PRL}. A comparison of $S_{xy}(E_F)$  calculated for  MBT$_\text{3SL}$/hBN and MBT$_\text{3SL}$ is shown in Fig.~\ref{fig:trilayer_ahc}. A good matching between the two can be observed, especially for the flat portion in the band gap where $S_{xy}=e^2/h$ for both, confirming insensitivity of the $C=1$ QAH state of MBT$_\text{3SL}$ to interfacing with hBN.

As we are seeking the high-$C$ QAH state realized at zero external field, the question arises whether it is supported by the magnetic ground state of $n$MBT$_\text{3SL}$/hBN, $n>1$. While we have shown above that the magnetic ordering inside MBT film is not affected by hBN, the interlayer exchange coupling between the local Mn moments through hBN should now be studied. Our total-energy calculations show that the AFM configuration is $\sim0.03\ \mathrm{meV}$ (per $2$ Mn atoms) lower in energy than the FM one (see Supplementary Note~SIII). Such a small energy difference is due to the large separation between neighboring Mn planes (about $17.8$ \AA{}) and vdW coupling between MBT and hBN. Although this number is at the limit of our calculation accuracy, assuming that its sign is correct yields a zero (non-zero) net magnetization in $n$MBT$_\text{3SL}$/hBN with even (odd) $n$. As the net $C$ is a sum of $C$'s of individual MBT$_\text{3SL}$ films, with the sign of $C$ changing with the magnetization direction, our result predicts that an $n$MBT$_\text{3SL}$/hBN heterostructure should have $C$ equal to $0$ ($1$) in its ground state. 

However, the weakness of the interlayer exchange coupling through hBN makes a reliable prediction of its sign hardly possible using the density functional theory, so it should be defined in future experiments. If the FM coupling through hBN occurs experimentally, the Chern numbers of the individual MBT films will sum up to create the high-$C$ state. However, even if the experiment would show the worst-case scenario of AFM coupling through hBN, a metamagnetic state that is appropriate for achieving the zero-field high-$C$ state in the $n$MBT$_\text{3SL}$/hBN ($n>1$) multilayers can nevertheless be "prepared" by the external magnetic field, in a manner we will describe below. In this state, the alignment of the MBT SLs across hBN would be FM, while the interlayer alignment inside MBT films stays AFM, thus guaranteeing an overall non-zero net magnetization increasing with $n$, in turn yielding the desired $C=n$ state. Crucial for the realization of such a state is a fact that the magnetic anisotropy energy of MBT$_\text{3SL}$ is $\sim0.22\ \mathrm{meV}$~\cite{Otrokov2019PRL}, several times larger than the energy difference of AFM and FM configurations across hBN. Note that the magnetic anisotropy energy of MBT$_\text{3SL}$ should not change appreciably upon its interfacing with hBN, as shown in Sec.~\ref{subsec:adsorption_study}.

We therefore propose the following procedure to induce the above described metamagnetic state. First, the external magnetic field of about $10$~\cite{Liu.nmat2020, Deng.sci2020, Ge.nsr2019} should be applied to overcome the AFM coupling inside individual MBT$_\text{3SL}$ films and polarize all SLs along the same direction. Next, the external magnetic field should be gradually reduced to zero leading to recovery of the uncompensated interlayer AFM state with non-zero magnetization in each individual MBT$_\text{3SL}$ film \cite{Otrokov2019PRL,Deng.sci2020}. However, their net magnetizations will remain parallel to each other even if the exchange coupling across hBN is AFM because the energy barrier due to anisotropy prevents the magnetization relaxation into the only slightly more favorable AFM configuration across hBN. Indeed, a similar metamagnetic state has been recently experimentally observed in bulk MnBi$_4$Te$_7$ and MnBi$_6$Te$_{10}$~\cite{Wu.sciadv2019, Vidal.prx2019, Hu.ncomms2020, Klimovskikh_npjQM2020, Tan.prl2020}, where the uniaxial anisotropy dominates over the AFM interlayer exchange coupling \cite{Tan.prl2020}. The magnetic anisotropy energy of MBT$_\text{3SL}$ is slightly larger, while the exchange coupling across hBN is significantly weaker than those in MnBi$_4$Te$_7$ and MnBi$_6$Te$_{10}$ (c.f.~Ref.~\cite{Klimovskikh_npjQM2020}), making the proposed realization of the metamagnetic state in $n$MBT$_\text{3SL}$/hBN feasible. 

We now have a firm basis to claim a possibility of the high-$C$ QAH state realization in $n$MBT$_\text{3SL}$/hBN, $n>1$. As in the preceding section, the high-$C$ state can be demonstrated by the $S_{xy}(E_F)$ and the edge electronic structure calculations, results of which is shown in Fig.~\ref{fig:trilayer_ahc}. Analogously to the $n$MBT$_\text{2SL}$/hBN multilayers, the $S_{xy}(E_F)$ for $n$MBT$_\text{3SL}$/hBN with $n=2$ shows a clear plateau within the band gap, where the conductance is equal to two conductance quanta, $2e^2/h$, i.e. $C=n=2$. Moreover, the edge electronic structure shows two edge modes traversing the band gap.
 
From these results it can be inferred that the $n$MBT$_\text{3SL}$/hBN, $n>1$, heterostructures have $C=n$ either in the metamagnetic state or even intrinsically, depending on the actual interlayer coupling over hBN. Similar results are expected for MBT/hBN multilayers based on MBT$_\text{5SL}$ and MBT$_\text{7SL}$ films.

\section{Discussion}

Using \emph{ab initio} and tight-binding calculations, we  studied novel \MBT/hBN multilayer heterostructures in which thin \MBT{} films are interlayed by hexagonal boron nitride monolayers. The van der Waals bonding between hBN and \MBT{} preserves the magnetic and electronic properties of the the latter, in particular, the $C=1$ Chern insulator state. Taking advantage of this, we showed that a stack of $n$ \MBT{} films with $C=1$ in the \MBT/hBN multilayer gives rise to a high Chern number state, $C=n$, characterized by $n$ chiral edge modes. 

There are two ways to achieve this state in the proposed heterostructures. The first way is to use the external magnetic field to drive \MBT{} films into a forced ferromagnetic state, which is nowadays widely used to observe a new kind of the quantum Hall effect that does not require formation of Landau levels \cite{Deng.sci2020, Liu.nmat2020, Ge.nsr2019, Cai.ncomms2022, Bai.arxiv2022, Hu.prm2021, Liu.ncomms2021, Ying.prb2022, Gao.nat2021, Ovchinnikov.nl2021}. In this case, both even and odd septuple layer \MBT{} films can be used. One may well expect the high Chern number state to persist up to the temperatures as high as $20-30$ K, as previously observed for the $C=1$ state in \MBT{} thin flakes \cite{Ge.nsr2019, Ying.prb2022}. The second way relies on use of the odd septuple layer films as they realize the quantum anomalous Hall state intrinsically \cite{Deng.sci2020}. Although a prior application of the external field might be needed to align the net magnetizations of the individual \MBT{} slabs (as it is done in the Cr-doped (Bi,Sb)$_2$Te$_3$ \cite{chang2013Science, Chang.nmat2015}), the high Chern number quantum anomalous Hall state can later be observed in remanence, i.e. at zero field. Currently, the observation temperature of the $C=1$ quantum anomalous Hall effect in \MBT{} is about $1.4$ K. Improving the structural quality of \MBT{} should allow the resolution of the intensely debated issue of its Dirac point gap size \cite{Otrokov2019Nature, Hao.prx2019, Li.prx2019, Chen.prx2019, Swatek.prb2020, Garnica.npjqm2022} and push the effect observation temperature towards \MBT's Néel point. The steps in this direction are currently underway~\cite{Hu.prm2021, Bai.arxiv2022, Liu.arxiv2022}.

\MBT{} thin films currently used in the quantized transport measurements are mostly obtained by exfoliation \cite{Deng.sci2020, Liu.nmat2020, Ge.nsr2019, Cai.ncomms2022, Hu.prm2021, Liu.ncomms2021, Ying.prb2022, Gao.nat2021, Ovchinnikov.nl2021}. Using exfoliation and transfer, which is a standard technique for construction of the van der Waals heterostructures \cite{Geim.nat2013, Novoselov.sci2016}, should be suitable to realize the here proposed \MBT/hBN multilayers as well. However, this approach will not necessarily favor formation of the interface structure with the lowest energy, but it could rather result in an arbitrary non-symmetric alignment between \MBT\, and hBN. Fortunately, our results show that the Chern insulator state of individual \MBT{} layers is not sensitive to the \MBT/hBN interface registry. Therefore the exfoliation and transfer strategy should be appropriate to synthesize the here proposed \MBT/hBN multilayers. The accumulated world-wide experience in the synthesis of two-dimensional van der Waals heterostructures is expected to greatly facilitate a prompt realization of the high Chern number state in the \MBT/hBN system.

In conclusion, we presented a novel concept of realization of a high Chern number magnetic topological insulator state, which relies on the use of fundamentally different, but highly compatible van der Waals materials, assembled in a multilayer heterostructure. While there is hardly a good alternative to hBN as an inert wide band gap monolayer insulator, other choices of the Chern insulators are in principle possible, such as transition metal dichalcogenide moiré superlattices \cite{Li.nature2021} or compounds of the \MBT\, family \cite{Otrokov.2dmat2017, Deng.nphys2021, Li.advmat2022, Wu.advmat2020}. The here proposed design allows a wide-range tuning of the Chern number, its upper limit being restricted only by the van der Waals heterostructures growth technology, making such multilayers interesting for future fundamental research and efficient interconnect technologies.

\section{Methods}

The first-principles calculations were carried out on the level of density functional theory (DFT) as implemented in the Vienna Ab-initio Simulation Package (VASP)~\cite{vasp1,vasp2,vasp3,vasp4,vasppaw}. All calculations shared several common parameters: the VASP's PAW datasets~\cite{vasppaw} were used, the plane wave cutoff was set to $500\ \mathrm{eV}$, the spin-orbit coupling (SOC) was enabled, the exchange-correlation functional was that of Perdew, Burke and Ernzerhof (PBE)~\cite{pbe}, while the vdW interaction was taken into the account through the Grimme D3 model with the Becke-Jonson cutoff~\cite{Grimme2010:JCP, Grimme2011:JCC}. The Mn $3d$-states were treated employing the GGA$+U$ approach \cite{Anisimov1991} within the Dudarev scheme \cite{Dudarev.prb1998}. The $U_\text{eff}=U-J$ value for the Mn 3$d$-states was chosen to be equal to 5.34~eV, as in previous works \cite{Otrokov2019Nature,Hao.prx2019}.

The DFT calculations were carried out with $11\times11\times1$ Monkhorst-Pack grid used to sample the Brillouin zone (BZ) for all but $n=5$ system, where $9\times9\times1$ Monkhorst-Pack grid was used instead, and the electronic convergence threshold was $10^{-6}\ \mathrm{eV}$. For structural relaxations, VASP's conjugate gradient algorithm was employed until the force on each atom decreased under $0.01\ \text{eV}$/\AA{}, and the electronic occupations were smeared by the Gaussian function of width of $0.01\ \text{eV}$. For the static calculations the tetrahedron method with Bl\"{o}chl corrections was used to treat the electronic occupations instead. The noncollinear intralayer AFM configuration was calculated using hexagonal cells containing three atoms per layer [$(\sqrt{3}\times\sqrt{3})R30^\circ$ in-plane periodicity] (see Supplementary Note SII A 3) and a $7\times 7\times 1$ BZ sampling.
For purpose of electronic structure visualization the band energies for selected systems were calculated along the $\overline{\mathrm{K}}-\overline{\Gamma}-\overline{\mathrm{M}}$ path by a non-self consistent DFT calculation, again with Gaussian smearing of $0.01\ \mathrm{meV}$. The magnetic anisotropy energy was calculated as explained in Ref.~\cite{Otrokov2019Nature}, using the $25\times25\times1$ Monkhorst-Pack grid and electronic convergence threshold of $10^{-8}\ \mathrm{eV}$.

The obtained Kohn-Sham functions were used to construct the maximally localized Wannier functions by the Wannier90 code~\cite{Mostofi2014CPC,Pizzi2020JOPCM}. We refer reader to the Supplementary Note.~SI for further details regarding the wannierization. The Wannier functions were in turn used to calculate anomalous Hall conductances of the proposed heterostructures through the Kubo formula as implemented in the WannierBerri code~\cite{tsirkin2021NJPCM}. In WannierBerri calculations the broadening of $10\ \mathrm{K}$ was used. For hBN-covered MBT$_\mathrm{2 SL}$ film, as well as MBT$_\mathrm{2 SL}$/hBN/MBT$_\mathrm{2 SL}$ heterostructure, the Chern number was also calculated by the Wilson loop method as implemented in Z2Pack~\cite{soluyanov2011prb,gresch2017prb}. The edge spectral functions were calculated by the Green function method~\cite{Sancho1985:JoPFMP} from the tight-binding Wannier Hamiltonian as implemented in WannierTools~\cite{Wu2018:CPC}. The number of principal layers in these calculations was set to four, while the details of the cell used can be found in the Supplementary Note SIV. 

\section*{Data availability statement}
Inputs and results are available from the corresponding authors upon a reasonable request.

\begin{acknowledgements}
The authors thank J.~Ibañez-Azpiroz, S.~S.~Tsirkin, I.~Souza, and M.~Garnica for stimulating discussions. M.B.~and M.M.O.~acknowledge support of the Spanish Ministerio de Ciencia e Innovacion (Grant no.~PID2019-103910GB-I00) and the University of the Basque Country (Grant no. IT1527-22). A.Yu.V.~and E.K.P.~acknowledge support of the Russian Science Foundation (grant \textnumero18-12-00169-p). E.V.C.~acknowledges support from Saint Petersburg State University (Grant ID 90383050). The calculations were performed using computational resources of Donostia International Physics Center (http://dipc.ehu.es/cc/) and Research Park of Saint Petersburg State University “Computing Center” (http://www.cc.spbu.ru/).
\end{acknowledgements}

\section*{Author Contributions}
The problem was conceptualized by MMO and EVC. Density functional theory calculations were performed by MB, AYV, EKP and MMO. Wannier90 and surface state (WannierTools) calculations were performed by MB and EKP. Z2Pack Chern number calculations were performed by MB and AYV. Hall conductance calculations were performed by MB. Analysis of results was done by MB and MMO. Figures were produced by MB.  The manuscript was written by MMO and MB, with input from other authors.

\section*{Competing interests}
Authors declare no competing interests.

\section*{Additional information}

\subsection*{Supplementary information}
Supporting Information acompanies the manuscript.


\begin{thebibliography}{10}

\bibitem{haldane1988PRL}
\bibinfo{author}{Haldane, F. D.~M.}
\newblock \bibinfo{title}{Model for a quantum {Hall} effect without {Landau}
  levels: Condensed-matter realization of the "parity anomaly"}.
\newblock \emph{\bibinfo{journal}{Phys. Rev. Lett.}}
  \textbf{\bibinfo{volume}{61}}, \bibinfo{pages}{2015--2018}
  (\bibinfo{year}{1988}).

\bibitem{He.nsr2014}
\bibinfo{author}{He, K.}, \bibinfo{author}{Wang, Y.} \& \bibinfo{author}{Xue,
  Q.-K.}
\newblock \bibinfo{title}{Quantum anomalous {Hall} effect}.
\newblock \emph{\bibinfo{journal}{Natl. Sci. Rev.}}
  \textbf{\bibinfo{volume}{1}}, \bibinfo{pages}{38--48} (\bibinfo{year}{2014}).

\bibitem{Tokura.nrp2019}
\bibinfo{author}{Tokura, Y.}, \bibinfo{author}{Yasuda, K.} \&
  \bibinfo{author}{Tsukazaki, A.}
\newblock \bibinfo{title}{Magnetic topological insulators}.
\newblock \emph{\bibinfo{journal}{Nat. Rev. Phys.}}
  \textbf{\bibinfo{volume}{1}}, \bibinfo{pages}{126--143}
  (\bibinfo{year}{2019}).

\bibitem{Chang.review2022}
\bibinfo{author}{Chang, C.-Z.}, \bibinfo{author}{Liu, C.-X.} \&
  \bibinfo{author}{MacDonald, A.~H.}
\newblock \bibinfo{title}{Colloquium: Quantum anomalous {Hall} effect}.
\newblock \emph{\bibinfo{journal}{arXiv:2202.13902}}  (\bibinfo{year}{2022}).

\bibitem{vonKlitzing.prl1980}
\bibinfo{author}{v.~Klitzing, K.}, \bibinfo{author}{Dorda, G.} \&
  \bibinfo{author}{Pepper, M.}
\newblock \bibinfo{title}{New method for high-accuracy determination of the
  fine-structure constant based on quantized {Hall} resistance}.
\newblock \emph{\bibinfo{journal}{Phys. Rev. Lett.}}
  \textbf{\bibinfo{volume}{45}}, \bibinfo{pages}{494--497}
  (\bibinfo{year}{1980}).

\bibitem{thouless1982PRL}
\bibinfo{author}{Thouless, D.~J.}, \bibinfo{author}{Kohmoto, M.},
  \bibinfo{author}{Nightingale, M.~P.} \& \bibinfo{author}{den Nijs, M.}
\newblock \bibinfo{title}{Quantized {Hall} conductance in a two-dimensional
  periodic potential}.
\newblock \emph{\bibinfo{journal}{Phys. Rev. Lett.}}
  \textbf{\bibinfo{volume}{49}}, \bibinfo{pages}{405} (\bibinfo{year}{1982}).

\bibitem{Zhang2012ISOP}
\bibinfo{author}{Zhang, X.} \& \bibinfo{author}{Zhang, S.-C.}
\newblock \bibinfo{title}{{Chiral interconnects based on topological
  insulators}}.
\newblock \emph{\bibinfo{journal}{Proc. of SPIE}}
  \textbf{\bibinfo{volume}{8373}}, \bibinfo{pages}{837309}
  (\bibinfo{year}{2012}).

\bibitem{Zhang.patent2016}
\bibinfo{author}{Zhang, S.-C.} \& \bibinfo{author}{Wang, J.}
\newblock \bibinfo{title}{Autobahn interconnect in {IC} with multiple
  conduction lanes} (\bibinfo{year}{2016}).
\newblock \bibinfo{note}{{U}S Patent App. 14/447,499}.

\bibitem{chang2013Science}
\bibinfo{author}{Chang, C.-Z.} \emph{et~al.}
\newblock \bibinfo{title}{Experimental observation of the quantum anomalous
  {Hall} effect in a magnetic topological insulator}.
\newblock \emph{\bibinfo{journal}{Science}} \textbf{\bibinfo{volume}{340}},
  \bibinfo{pages}{167--170} (\bibinfo{year}{2013}).

\bibitem{Kou.prl2014}
\bibinfo{author}{Kou, X.} \emph{et~al.}
\newblock \bibinfo{title}{Scale-invariant quantum anomalous {Hall} effect in
  magnetic topological insulators beyond the two-dimensional limit}.
\newblock \emph{\bibinfo{journal}{Phys. Rev. Lett.}}
  \textbf{\bibinfo{volume}{113}}, \bibinfo{pages}{137201}
  (\bibinfo{year}{2014}).

\bibitem{Kandala.ncomms2015}
\bibinfo{author}{Kandala, A.}, \bibinfo{author}{Richardella, A.},
  \bibinfo{author}{Kempinger, S.}, \bibinfo{author}{Liu, C.-X.} \&
  \bibinfo{author}{Samarth, N.}
\newblock \bibinfo{title}{Giant anisotropic magnetoresistance in a quantum
  anomalous {Hall} insulator}.
\newblock \emph{\bibinfo{journal}{Nat. Commun.}} \textbf{\bibinfo{volume}{6}},
  \bibinfo{pages}{7434} (\bibinfo{year}{2015}).

\bibitem{Mogi.apl2015}
\bibinfo{author}{Mogi, M.} \emph{et~al.}
\newblock \bibinfo{title}{Magnetic modulation doping in topological insulators
  toward higher-temperature quantum anomalous {Hall} effect}.
\newblock \emph{\bibinfo{journal}{Appl. Phys. Lett.}}
  \textbf{\bibinfo{volume}{107}} (\bibinfo{year}{2015}).

\bibitem{Gotz.apl2018}
\bibinfo{author}{G{\"o}tz, M.} \emph{et~al.}
\newblock \bibinfo{title}{Precision measurement of the quantized anomalous
  {Hall} resistance at zero magnetic field}.
\newblock \emph{\bibinfo{journal}{Appl. Phys. Lett.}}
  \textbf{\bibinfo{volume}{112}}, \bibinfo{pages}{072102}
  (\bibinfo{year}{2018}).

\bibitem{Okazaki.nphys2022}
\bibinfo{author}{Okazaki, Y.} \emph{et~al.}
\newblock \bibinfo{title}{Quantum anomalous {Hall} effect with a permanent
  magnet defines a quantum resistance standard}.
\newblock \emph{\bibinfo{journal}{Nat. Phys.}} \textbf{\bibinfo{volume}{18}},
  \bibinfo{pages}{25--29} (\bibinfo{year}{2022}).

\bibitem{wang2013PRL}
\bibinfo{author}{Wang, J.}, \bibinfo{author}{Lian, B.}, \bibinfo{author}{Zhang,
  H.}, \bibinfo{author}{Xu, Y.} \& \bibinfo{author}{Zhang, S.-C.}
\newblock \bibinfo{title}{Quantum anomalous {Hall} effect with higher
  plateaus}.
\newblock \emph{\bibinfo{journal}{Phys. Rev. Lett.}}
  \textbf{\bibinfo{volume}{111}}, \bibinfo{pages}{136801}
  (\bibinfo{year}{2013}).

\bibitem{Jiang.cpl2018}
\bibinfo{author}{Jiang, G.} \emph{et~al.}
\newblock \bibinfo{title}{Quantum anomalous {Hall} multilayers grown by
  molecular beam epitaxy}.
\newblock \emph{\bibinfo{journal}{Chin. Phys. Lett.}}
  \textbf{\bibinfo{volume}{35}}, \bibinfo{pages}{076802}
  (\bibinfo{year}{2018}).

\bibitem{zhao2020nature}
\bibinfo{author}{Zhao, Y.-F.} \emph{et~al.}
\newblock \bibinfo{title}{Tuning the {Chern} number in quantum anomalous {Hall}
  insulators}.
\newblock \emph{\bibinfo{journal}{Nature}} \textbf{\bibinfo{volume}{588}},
  \bibinfo{pages}{419--423} (\bibinfo{year}{2020}).

\bibitem{lee2015pnasu}
\bibinfo{author}{Lee, C.} \emph{et~al.}
\newblock \bibinfo{title}{Imaging {Dirac}-mass disorder from magnetic dopant
  atoms in the ferromagnetic topological insulator}.
\newblock \emph{\bibinfo{journal}{Proc. Natl. Acad. Sci. USA}}
  \textbf{\bibinfo{volume}{112}}, \bibinfo{pages}{1316} (\bibinfo{year}{2015}).

\bibitem{Chong.nl2020}
\bibinfo{author}{Chong, Y.~X.} \emph{et~al.}
\newblock \bibinfo{title}{Severe {Dirac} mass gap suppression in
  {Sb}$_2${Te}$_3$-based quantum anomalous {Hall} materials}.
\newblock \emph{\bibinfo{journal}{Nano Lett.}} \textbf{\bibinfo{volume}{20}},
  \bibinfo{pages}{8001--8007} (\bibinfo{year}{2020}).

\bibitem{lachman2015sciadv}
\bibinfo{author}{Lachman, E.~O.} \emph{et~al.}
\newblock \bibinfo{title}{Visualization of superparamagnetic dynamics in
  magnetic topological insulators}.
\newblock \emph{\bibinfo{journal}{Sci. Adv.}} \textbf{\bibinfo{volume}{1}},
  \bibinfo{pages}{e1500740} (\bibinfo{year}{2015}).

\bibitem{Ou.advmat2018}
\bibinfo{author}{Ou, Y.} \emph{et~al.}
\newblock \bibinfo{title}{Enhancing the quantum anomalous {Hall} effect by
  magnetic codoping in a topological insulator}.
\newblock \emph{\bibinfo{journal}{Adv. Mater.}} \textbf{\bibinfo{volume}{30}},
  \bibinfo{pages}{1703062} (\bibinfo{year}{2018}).

\bibitem{Otrokov2019Nature}
\bibinfo{author}{Otrokov, M.~M.} \emph{et~al.}
\newblock \bibinfo{title}{Prediction and observation of an antiferromagnetic
  topological insulator}.
\newblock \emph{\bibinfo{journal}{Nature}} \textbf{\bibinfo{volume}{576}},
  \bibinfo{pages}{416--422} (\bibinfo{year}{2019}).

\bibitem{Deng.sci2020}
\bibinfo{author}{Deng, Y.} \emph{et~al.}
\newblock \bibinfo{title}{Quantum anomalous {Hall} effect in intrinsic magnetic
  topological insulator {MnBi}$_2${Te}$_4$}.
\newblock \emph{\bibinfo{journal}{Science}} \textbf{\bibinfo{volume}{367}},
  \bibinfo{pages}{895--900} (\bibinfo{year}{2020}).

\bibitem{Deng.nphys2021}
\bibinfo{author}{Deng, H.} \emph{et~al.}
\newblock \bibinfo{title}{High-temperature quantum anomalous {Hall} regime in a
  {MnBi}$_2${Te}$_4$/{Bi}$_2${Te}$_3$ superlattice}.
\newblock \emph{\bibinfo{journal}{Nat. Phys.}} \textbf{\bibinfo{volume}{17}},
  \bibinfo{pages}{36–42} (\bibinfo{year}{2021}).

\bibitem{Serlin.sci2020}
\bibinfo{author}{Serlin, M.} \emph{et~al.}
\newblock \bibinfo{title}{Intrinsic quantized anomalous {Hall} effect in a
  moir{\'e} heterostructure}.
\newblock \emph{\bibinfo{journal}{Science}} \textbf{\bibinfo{volume}{367}},
  \bibinfo{pages}{900--903} (\bibinfo{year}{2020}).

\bibitem{Li.nature2021}
\bibinfo{author}{Li, T.} \emph{et~al.}
\newblock \bibinfo{title}{Quantum anomalous {Hall} effect from intertwined
  moir{\'e} bands}.
\newblock \emph{\bibinfo{journal}{Nature}} \textbf{\bibinfo{volume}{600}},
  \bibinfo{pages}{641--646} (\bibinfo{year}{2021}).

\bibitem{Otrokov2019PRL}
\bibinfo{author}{Otrokov, M.~M.} \emph{et~al.}
\newblock \bibinfo{title}{Unique thickness-dependent properties of the van der
  {Waals} interlayer antiferromagnet {MnBi}$_{2}${Te}$_{4}$ films}.
\newblock \emph{\bibinfo{journal}{Phys. Rev. Lett.}}
  \textbf{\bibinfo{volume}{122}}, \bibinfo{pages}{107202}
  (\bibinfo{year}{2019}).

\bibitem{Mong.prb2010}
\bibinfo{author}{Mong, R. S.~K.}, \bibinfo{author}{Essin, A.~M.} \&
  \bibinfo{author}{Moore, J.~E.}
\newblock \bibinfo{title}{Antiferromagnetic topological insulators}.
\newblock \emph{\bibinfo{journal}{Phys. Rev. B}} \textbf{\bibinfo{volume}{81}},
  \bibinfo{pages}{245209} (\bibinfo{year}{2010}).

\bibitem{Li.sciadv2019}
\bibinfo{author}{Li, J.} \emph{et~al.}
\newblock \bibinfo{title}{Intrinsic magnetic topological insulators in van der
  {Waals} layered {MnBi}$_2${Te}$_4$-family materials}.
\newblock \emph{\bibinfo{journal}{Sci. Adv.}} \textbf{\bibinfo{volume}{5}},
  \bibinfo{pages}{eaaw5685} (\bibinfo{year}{2019}).

\bibitem{Zhang.prl2019}
\bibinfo{author}{Zhang, D.} \emph{et~al.}
\newblock \bibinfo{title}{Topological axion states in the magnetic insulator
  {MnBi}$_{2}${Te}$_{4}$ with the quantized magnetoelectric effect}.
\newblock \emph{\bibinfo{journal}{Phys. Rev. Lett.}}
  \textbf{\bibinfo{volume}{122}}, \bibinfo{pages}{206401}
  (\bibinfo{year}{2019}).

\bibitem{Yan.prm2019}
\bibinfo{author}{Yan, J.-Q.} \emph{et~al.}
\newblock \bibinfo{title}{Crystal growth and magnetic structure of
  {MnBi}$_{2}${Te}$_{4}$}.
\newblock \emph{\bibinfo{journal}{Phys. Rev. Mater.}}
  \textbf{\bibinfo{volume}{3}}, \bibinfo{pages}{064202} (\bibinfo{year}{2019}).

\bibitem{Liu.nmat2020}
\bibinfo{author}{Liu, C.} \emph{et~al.}
\newblock \bibinfo{title}{Robust axion insulator and {Chern} insulator phases
  in a two-dimensional antiferromagnetic topological insulator}.
\newblock \emph{\bibinfo{journal}{Nat. Mater.}} \textbf{\bibinfo{volume}{19}},
  \bibinfo{pages}{522--527} (\bibinfo{year}{2020}).

\bibitem{Ge.nsr2019}
\bibinfo{author}{Ge, J.} \emph{et~al.}
\newblock \bibinfo{title}{{{High}-{Chern}-number and high-temperature quantum
  {Hall} effect without {Landau} levels}}.
\newblock \emph{\bibinfo{journal}{National Science Review}}
  \textbf{\bibinfo{volume}{7}}, \bibinfo{pages}{1280--1287}
  (\bibinfo{year}{2020}).

\bibitem{Cai.ncomms2022}
\bibinfo{author}{Cai, J.} \emph{et~al.}
\newblock \bibinfo{title}{Electric control of a canted-antiferromagnetic
  {Chern} insulator}.
\newblock \emph{\bibinfo{journal}{Nat. Commun.}} \textbf{\bibinfo{volume}{13}},
  \bibinfo{pages}{1668--1675} (\bibinfo{year}{2022}).

\bibitem{Hu.prm2021}
\bibinfo{author}{Hu, C.} \emph{et~al.}
\newblock \bibinfo{title}{Growth, characterization, and {Chern} insulator state
  in {MnBi}$_2${Te}$_4$ via the chemical vapor transport method}.
\newblock \emph{\bibinfo{journal}{Phys. Rev. Mater.}}
  \textbf{\bibinfo{volume}{5}}, \bibinfo{pages}{124206} (\bibinfo{year}{2021}).

\bibitem{Liu.ncomms2021}
\bibinfo{author}{Liu, C.} \emph{et~al.}
\newblock \bibinfo{title}{Magnetic-field-induced robust zero {Hall} plateau
  state in {MnBi}$_2${Te}$_4$ chern insulator}.
\newblock \emph{\bibinfo{journal}{Nat. Commun.}} \textbf{\bibinfo{volume}{12}},
  \bibinfo{pages}{4647--4653} (\bibinfo{year}{2021}).

\bibitem{Ying.prb2022}
\bibinfo{author}{Ying, Z.} \emph{et~al.}
\newblock \bibinfo{title}{Experimental evidence for dissipationless transport
  of the chiral edge state of the high-field {Chern} insulator in
  {MnBi}$_{2}${Te}$_{4}$ nanodevices}.
\newblock \emph{\bibinfo{journal}{Phys. Rev. B}}
  \textbf{\bibinfo{volume}{105}}, \bibinfo{pages}{085412}
  (\bibinfo{year}{2022}).

\bibitem{Gao.nat2021}
\bibinfo{author}{Gao, A.} \emph{et~al.}
\newblock \bibinfo{title}{Layer {Hall} effect in a {2D} topological axion
  antiferromagnet}.
\newblock \emph{\bibinfo{journal}{Nature}} \textbf{\bibinfo{volume}{595}},
  \bibinfo{pages}{521--525} (\bibinfo{year}{2021}).

\bibitem{Ovchinnikov.nl2021}
\bibinfo{author}{Ovchinnikov, D.} \emph{et~al.}
\newblock \bibinfo{title}{Intertwined topological and magnetic orders in
  atomically thin {Chern} insulator {MnBi}$_2${Te}$_4$}.
\newblock \emph{\bibinfo{journal}{Nano Lett.}} \textbf{\bibinfo{volume}{21}},
  \bibinfo{pages}{2544--2550} (\bibinfo{year}{2021}).

\bibitem{Britnell2012Science}
\bibinfo{author}{Britnell, L.} \emph{et~al.}
\newblock \bibinfo{title}{Field-effect tunneling transistor based on vertical
  graphene heterostructures}.
\newblock \emph{\bibinfo{journal}{Science}} \textbf{\bibinfo{volume}{335}},
  \bibinfo{pages}{947--950} (\bibinfo{year}{2012}).

\bibitem{Dean2010NatNano}
\bibinfo{author}{Dean, C.~R.} \emph{et~al.}
\newblock \bibinfo{title}{Boron nitride substrates for high-quality graphene
  electronics}.
\newblock \emph{\bibinfo{journal}{Nat. Nanotechnol.}}
  \textbf{\bibinfo{volume}{5}}, \bibinfo{pages}{722--726}
  (\bibinfo{year}{2010}).

\bibitem{Geim.nat2013}
\bibinfo{author}{Geim, A.~K.} \& \bibinfo{author}{Grigorieva, I.~V.}
\newblock \bibinfo{title}{Van der {Waals} heterostructures}.
\newblock \emph{\bibinfo{journal}{Nature}} \textbf{\bibinfo{volume}{499}},
  \bibinfo{pages}{419--425} (\bibinfo{year}{2013}).

\bibitem{Novoselov.sci2016}
\bibinfo{author}{Novoselov, K.}, \bibinfo{author}{Mishchenko, A.},
  \bibinfo{author}{Carvalho, A.} \& \bibinfo{author}{Castro~Neto, A.}
\newblock \bibinfo{title}{{2D} materials and van der {Waals} heterostructures}.
\newblock \emph{\bibinfo{journal}{Science}} \textbf{\bibinfo{volume}{353}},
  \bibinfo{pages}{aac9439} (\bibinfo{year}{2016}).

\bibitem{Lee.cec2013}
\bibinfo{author}{Lee, D.~S.} \emph{et~al.}
\newblock \bibinfo{title}{Crystal structure, properties and nanostructuring of
  a new layered chalcogenide semiconductor, {Bi}$_2${MnTe}$_4$}.
\newblock \emph{\bibinfo{journal}{Cryst. Eng. Comm.}}
  \textbf{\bibinfo{volume}{15}}, \bibinfo{pages}{5532--5538}
  (\bibinfo{year}{2013}).

\bibitem{Zeugner.cm2019}
\bibinfo{author}{Zeugner, A.} \emph{et~al.}
\newblock \bibinfo{title}{Chemical aspects of the candidate antiferromagnetic
  topological insulator {MnBi}$_2${Te}$_4$}.
\newblock \emph{\bibinfo{journal}{Chem. Mater.}} \textbf{\bibinfo{volume}{31}},
  \bibinfo{pages}{2795--2806} (\bibinfo{year}{2019}).

\bibitem{Lee.prr2019}
\bibinfo{author}{Lee, S.~H.} \emph{et~al.}
\newblock \bibinfo{title}{Spin scattering and noncollinear spin
  structure-induced intrinsic anomalous {Hall} effect in antiferromagnetic
  topological insulator {MnBi}$_{2}${Te}$_{4}$}.
\newblock \emph{\bibinfo{journal}{Phys. Rev. Res.}}
  \textbf{\bibinfo{volume}{1}}, \bibinfo{pages}{012011} (\bibinfo{year}{2019}).

\bibitem{Gao2021PRM}
\bibinfo{author}{Gao, R.}, \bibinfo{author}{Qin, G.}, \bibinfo{author}{Qi, S.},
  \bibinfo{author}{Qiao, Z.} \& \bibinfo{author}{Ren, W.}
\newblock \bibinfo{title}{Quantum anomalous {Hall} effect in
  {MnBi}$_{2}${Te}$_{4}$ van der {Waals} heterostructures}.
\newblock \emph{\bibinfo{journal}{Phys. Rev. Mater.}}
  \textbf{\bibinfo{volume}{5}}, \bibinfo{pages}{114201} (\bibinfo{year}{2021}).

\bibitem{Bai.arxiv2022}
\bibinfo{author}{Bai, Y.} \emph{et~al.}
\newblock \bibinfo{title}{Quantized anomalous {Hall} resistivity achieved in
  molecular beam epitaxy-grown {MnBi}$_2${Te}$_4$ thin films}.
\newblock \emph{\bibinfo{journal}{arXiv preprint arXiv:2206.03773}}
  (\bibinfo{year}{2022}).

\bibitem{lujan2022:natcomm}
\bibinfo{author}{Lujan, D.} \emph{et~al.}
\newblock \bibinfo{title}{Magnons and magnetic fluctuations in atomically thin
  {MnBi}$_2${Te}$_4$}.
\newblock \emph{\bibinfo{journal}{Nature Communications}}
  \textbf{\bibinfo{volume}{13}}, \bibinfo{pages}{2527--2534}
  (\bibinfo{year}{2022}).

\bibitem{Wu.sciadv2019}
\bibinfo{author}{Wu, J.} \emph{et~al.}
\newblock \bibinfo{title}{Natural van der {Waals} heterostructural single
  crystals with both magnetic and topological properties}.
\newblock \emph{\bibinfo{journal}{Sci. Adv.}} \textbf{\bibinfo{volume}{5}},
  \bibinfo{pages}{eaax9989} (\bibinfo{year}{2019}).

\bibitem{Vidal.prx2019}
\bibinfo{author}{Vidal, R.~C.} \emph{et~al.}
\newblock \bibinfo{title}{Topological electronic structure and intrinsic
  magnetization in {MnBi}$_4${Te}$_7$: A {Bi}$_2${Te}$_3$ derivative with a
  periodic {Mn} sublattice}.
\newblock \emph{\bibinfo{journal}{Phys. Rev. X}} \textbf{\bibinfo{volume}{9}},
  \bibinfo{pages}{041065} (\bibinfo{year}{2019}).

\bibitem{Hu.ncomms2020}
\bibinfo{author}{Hu, C.} \emph{et~al.}
\newblock \bibinfo{title}{A van der {Waals} antiferromagnetic topological
  insulator with weak interlayer magnetic coupling}.
\newblock \emph{\bibinfo{journal}{Nat. Commun.}} \textbf{\bibinfo{volume}{11}},
  \bibinfo{pages}{97--105} (\bibinfo{year}{2020}).

\bibitem{Klimovskikh_npjQM2020}
\bibinfo{author}{Klimovskikh, I.~I.} \emph{et~al.}
\newblock \bibinfo{title}{Tunable {3D}/{2D} magnetism in the
  ({MnBi}$_2${Te}$_4$)({Bi}$_2${Te}$_3$)$_m$ topological insulators family}.
\newblock \emph{\bibinfo{journal}{npj Quantum Mater.}}
  \textbf{\bibinfo{volume}{5}}, \bibinfo{pages}{9} (\bibinfo{year}{2020}).

\bibitem{Tan.prl2020}
\bibinfo{author}{Tan, A.} \emph{et~al.}
\newblock \bibinfo{title}{Metamagnetism of weakly coupled antiferromagnetic
  topological insulators}.
\newblock \emph{\bibinfo{journal}{Phys. Rev. Lett.}}
  \textbf{\bibinfo{volume}{124}}, \bibinfo{pages}{197201}
  (\bibinfo{year}{2020}).
\newblock

\bibitem{Chang.nmat2015}
\bibinfo{author}{Chang, C.-Z.} \emph{et~al.}
\newblock \bibinfo{title}{High-precision realization of robust quantum
  anomalous {Hall} state in a hard ferromagnetic topological insulator}.
\newblock \emph{\bibinfo{journal}{Nat. Mater.}} \textbf{\bibinfo{volume}{14}},
  \bibinfo{pages}{473--477} (\bibinfo{year}{2015}).

\bibitem{Hao.prx2019}
\bibinfo{author}{Hao, Y.-J.} \emph{et~al.}
\newblock \bibinfo{title}{Gapless surface {Dirac} cone in antiferromagnetic
  topological insulator {MnBi}$_2${Te}$_{4}$}.
\newblock \emph{\bibinfo{journal}{Phys. Rev. X}} \textbf{\bibinfo{volume}{9}},
  \bibinfo{pages}{041038} (\bibinfo{year}{2019}).

\bibitem{Li.prx2019}
\bibinfo{author}{Li, H.} \emph{et~al.}
\newblock \bibinfo{title}{Dirac surface states in intrinsic magnetic
  topological insulators {EuSn}$_{2}${As}$_{2}$ and
  {MnBi}$_{2n}${Te}$_{3n+1}$}.
\newblock \emph{\bibinfo{journal}{Phys. Rev. X}} \textbf{\bibinfo{volume}{9}},
  \bibinfo{pages}{041039} (\bibinfo{year}{2019}).

\bibitem{Chen.prx2019}
\bibinfo{author}{Chen, Y.~J.} \emph{et~al.}
\newblock \bibinfo{title}{Topological electronic structure and its temperature
  evolution in antiferromagnetic topological insulator {MnBi}$_{2}${Te}$_{4}$}.
\newblock \emph{\bibinfo{journal}{Phys. Rev. X}} \textbf{\bibinfo{volume}{9}},
  \bibinfo{pages}{041040} (\bibinfo{year}{2019}).

\bibitem{Swatek.prb2020}
\bibinfo{author}{Swatek, P.} \emph{et~al.}
\newblock \bibinfo{title}{Gapless {Dirac} surface states in the
  antiferromagnetic topological insulator {MnBi}$_{2}${Te}$_{4}$}.
\newblock \emph{\bibinfo{journal}{Phys. Rev. B}}
  \textbf{\bibinfo{volume}{101}}, \bibinfo{pages}{161109}
  (\bibinfo{year}{2020}).

\bibitem{Garnica.npjqm2022}
\bibinfo{author}{Garnica, M.} \emph{et~al.}
\newblock \bibinfo{title}{Native point defects and their implications for the
  {Dirac} point gap at {MnBi}$_2${Te}$_4$ (0001)}.
\newblock \emph{\bibinfo{journal}{npj Quantum Mater.}}
  \textbf{\bibinfo{volume}{7}}, \bibinfo{pages}{7} (\bibinfo{year}{2022}).

\bibitem{Liu.arxiv2022}
\bibinfo{author}{Liu, M.} \emph{et~al.}
\newblock \bibinfo{title}{Visualizing the interplay of {Dirac} mass gap and
  magnetism at nanoscale in intrinsic magnetic topological insulators}.
\newblock \emph{\bibinfo{journal}{arXiv preprint arXiv:2205.09195}}
  (\bibinfo{year}{2022}).

\bibitem{Otrokov.2dmat2017}
\bibinfo{author}{Otrokov, M.~M.} \emph{et~al.}
\newblock \bibinfo{title}{Highly-ordered wide bandgap materials for quantized
  anomalous {Hall} and magnetoelectric effects}.
\newblock \emph{\bibinfo{journal}{2D Mater.}} \textbf{\bibinfo{volume}{4}},
  \bibinfo{pages}{025082} (\bibinfo{year}{2017}).

\bibitem{Li.advmat2022}
\bibinfo{author}{Li, Q.} \emph{et~al.}
\newblock \bibinfo{title}{Large magnetic gap in a designer
  ferromagnet--topological insulator--ferromagnet heterostructure}.
\newblock \emph{\bibinfo{journal}{Adv. Mater.}} \textbf{\bibinfo{volume}{34}},
  \bibinfo{pages}{2107520} (\bibinfo{year}{2022}).

\bibitem{Wu.advmat2020}
\bibinfo{author}{Wu, J.} \emph{et~al.}
\newblock \bibinfo{title}{Toward {2D} magnets in the
  {MnBi}$_2${Te}$_4$)({Bi}$_2${Te}$_3$)$_n$ bulk crystal}.
\newblock \emph{\bibinfo{journal}{Adv. Mater.}} \textbf{\bibinfo{volume}{32}},
  \bibinfo{pages}{2001815} (\bibinfo{year}{2020}).

\bibitem{vasp1}
\bibinfo{author}{Kresse, G.} \& \bibinfo{author}{Hafner, J.}
\newblock \bibinfo{title}{Ab initio molecular dynamics for liquid metals}.
\newblock \emph{\bibinfo{journal}{Phys. Rev. B}} \textbf{\bibinfo{volume}{47}},
  \bibinfo{pages}{558--561} (\bibinfo{year}{1993}).

\bibitem{vasp2}
\bibinfo{author}{Kresse, G.} \& \bibinfo{author}{Hafner, J.}
\newblock \bibinfo{title}{Ab initio molecular-dynamics simulation of the
  liquid-metal--amorphous-semiconductor transition in germanium}.
\newblock \emph{\bibinfo{journal}{Phys. Rev. B}} \textbf{\bibinfo{volume}{49}},
  \bibinfo{pages}{14251--14269} (\bibinfo{year}{1994}).

\bibitem{vasp3}
\bibinfo{author}{Kresse, G.} \& \bibinfo{author}{Furthmüller, J.}
\newblock \bibinfo{title}{Efficiency of ab-initio total energy calculations for
  metals and semiconductors using a plane-wave basis set}.
\newblock \emph{\bibinfo{journal}{Comput. Mater. Sci.}}
  \textbf{\bibinfo{volume}{6}}, \bibinfo{pages}{15 -- 50}
  (\bibinfo{year}{1996}).

\bibitem{vasp4}
\bibinfo{author}{Kresse, G.} \& \bibinfo{author}{Furthm\"uller, J.}
\newblock \bibinfo{title}{Efficient iterative schemes for ab initio
  total-energy calculations using a plane-wave basis set}.
\newblock \emph{\bibinfo{journal}{Phys. Rev. B}} \textbf{\bibinfo{volume}{54}},
  \bibinfo{pages}{11169--11186} (\bibinfo{year}{1996}).

\bibitem{vasppaw}
\bibinfo{author}{Kresse, G.} \& \bibinfo{author}{Joubert, D.}
\newblock \bibinfo{title}{From ultrasoft pseudopotentials to the projector
  augmented-wave method}.
\newblock \emph{\bibinfo{journal}{Phys. Rev. B}} \textbf{\bibinfo{volume}{59}},
  \bibinfo{pages}{1758--1775} (\bibinfo{year}{1999}).

\bibitem{pbe}
\bibinfo{author}{Perdew, J.~P.}, \bibinfo{author}{Burke, K.} \&
  \bibinfo{author}{Ernzerhof, M.}
\newblock \bibinfo{title}{Generalized gradient approximation made simple}.
\newblock \emph{\bibinfo{journal}{Phys. Rev. Lett.}}
  \textbf{\bibinfo{volume}{77}}, \bibinfo{pages}{3865} (\bibinfo{year}{1996}).

\bibitem{Grimme2010:JCP}
\bibinfo{author}{Grimme, S.}, \bibinfo{author}{Antony, J.},
  \bibinfo{author}{Ehrlich, S.} \& \bibinfo{author}{Krieg, H.}
\newblock \bibinfo{title}{A consistent and accurate ab initio parametrization
  of density functional dispersion correction ({DFT-D}) for the 94 elements
  {H-Pu}}.
\newblock \emph{\bibinfo{journal}{J. Chem. Phys.}}
  \textbf{\bibinfo{volume}{132}}, \bibinfo{pages}{154104}
  (\bibinfo{year}{2010}).

\bibitem{Grimme2011:JCC}
\bibinfo{author}{Grimme, S.}, \bibinfo{author}{Ehrlich, S.} \&
  \bibinfo{author}{Goerigk, L.}
\newblock \bibinfo{title}{Effect of the damping function in dispersion
  corrected density functional theory}.
\newblock \emph{\bibinfo{journal}{J. Comput. Chem.}}
  \textbf{\bibinfo{volume}{32}}, \bibinfo{pages}{1456--1465}
  (\bibinfo{year}{2011}).

\bibitem{Anisimov1991}
\bibinfo{author}{Anisimov, V.~I.}, \bibinfo{author}{Zaanen, J.} \&
  \bibinfo{author}{Andersen, O.~K.}
\newblock \bibinfo{title}{Band theory and {Mott} insulators: {Hubbard} {U}
  instead of {Stoner} {I}}.
\newblock \emph{\bibinfo{journal}{Phys. Rev. B}} \textbf{\bibinfo{volume}{44}},
  \bibinfo{pages}{943--954} (\bibinfo{year}{1991}).

\bibitem{Dudarev.prb1998}
\bibinfo{author}{Dudarev, S.~L.}, \bibinfo{author}{Botton, G.~A.},
  \bibinfo{author}{Savrasov, S.~Y.}, \bibinfo{author}{Humphreys, C.~J.} \&
  \bibinfo{author}{Sutton, A.~P.}
\newblock \bibinfo{title}{Electron-energy-loss spectra and the structural
  stability of nickel oxide: An {LSDA+U} study}.
\newblock \emph{\bibinfo{journal}{Phys. Rev. B}} \textbf{\bibinfo{volume}{57}},
  \bibinfo{pages}{1505--1509} (\bibinfo{year}{1998}).

\bibitem{Mostofi2014CPC}
\bibinfo{author}{Mostofi, A.~A.} \emph{et~al.}
\newblock \bibinfo{title}{An updated version of {Wannier}90: A tool for
  obtaining maximally-localised {Wannier} functions}.
\newblock \emph{\bibinfo{journal}{Comput. Phys. Commun.}}
  \textbf{\bibinfo{volume}{185}}, \bibinfo{pages}{2309--2310}
  (\bibinfo{year}{2014}).

\bibitem{Pizzi2020JOPCM}
\bibinfo{author}{Pizzi, G.} \emph{et~al.}
\newblock \bibinfo{title}{Wannier90 as a community code: new features and
  applications}.
\newblock \emph{\bibinfo{journal}{J. Condens. Matter Phys.}}
  \textbf{\bibinfo{volume}{32}}, \bibinfo{pages}{165902}
  (\bibinfo{year}{2020}).

\bibitem{tsirkin2021NJPCM}
\bibinfo{author}{Tsirkin, S.~S.}
\newblock \bibinfo{title}{High performance {Wannier} interpolation of {Berry}
  curvature and related quantities with {WannierBerri} code}.
\newblock \emph{\bibinfo{journal}{Npj Comput. Mater.}}
  \textbf{\bibinfo{volume}{7}}, \bibinfo{pages}{1--9} (\bibinfo{year}{2021}).

\bibitem{soluyanov2011prb}
\bibinfo{author}{Soluyanov, A.~A.} \& \bibinfo{author}{Vanderbilt, D.}
\newblock \bibinfo{title}{Computing topological invariants without inversion
  symmetry}.
\newblock \emph{\bibinfo{journal}{Phys. Rev. B}} \textbf{\bibinfo{volume}{83}},
  \bibinfo{pages}{235401} (\bibinfo{year}{2011}).

\bibitem{gresch2017prb}
\bibinfo{author}{Gresch, D.} \emph{et~al.}
\newblock \bibinfo{title}{{Z2Pack}: Numerical implementation of hybrid
  {Wannier} centers for identifying topological materials}.
\newblock \emph{\bibinfo{journal}{Phys. Rev. B}} \textbf{\bibinfo{volume}{95}},
  \bibinfo{pages}{075146} (\bibinfo{year}{2017}).

\bibitem{Sancho1985:JoPFMP}
\bibinfo{author}{Sancho, M. P.~L.}, \bibinfo{author}{Sancho, J. M.~L.},
  \bibinfo{author}{Sancho, J. M.~L.} \& \bibinfo{author}{Rubio, J.}
\newblock \bibinfo{title}{Highly convergent schemes for the calculation of bulk
  and surface {Green} functions}.
\newblock \emph{\bibinfo{journal}{Journal of Physics F: Metal Physics}}
  \textbf{\bibinfo{volume}{15}}, \bibinfo{pages}{851--858}
  (\bibinfo{year}{1985}).

\bibitem{Wu2018:CPC}
\bibinfo{author}{Wu, Q.}, \bibinfo{author}{Zhang, S.}, \bibinfo{author}{Song,
  H.-F.}, \bibinfo{author}{Troyer, M.} \& \bibinfo{author}{Soluyanov, A.~A.}
\newblock \bibinfo{title}{{WannierTools} : An open-source software package for
  novel topological materials}.
\newblock \emph{\bibinfo{journal}{Computer Physics Communications}}
  \textbf{\bibinfo{volume}{224}}, \bibinfo{pages}{405 -- 416}
  (\bibinfo{year}{2018}).

\end{thebibliography}
\end{document}